\documentclass
[preprint,pre,nofootinbib,showpacs,titlepage,twocolumn,tightenlines,10pt]{revtex4}%
\usepackage{amsfonts}
\usepackage{amsmath}
\usepackage{amssymb}
\usepackage{graphicx}
\usepackage{acronym}%
\newtheorem{theorem}{Theorem}
\newtheorem{acknowledgement}[theorem]{Acknowledgement}

\begin{document}
\title{Laplacian Spectra as a Diagnostic Tool for Network Structure and Dynamics.}
\author{Patrick N. McGraw and Michael Menzinger}
\affiliation{Department of Chemistry, University of Toronto, Toronto, Ontario, Canada M5S 3H6}

\begin{abstract}
\ We examine numerically the three-way relationships among structure,
Laplacian spectra and frequency synchronization dynamics on complex networks.
\ We study the effects of clustering, degree distribution and a particular
type of coupling asymmetry (input normalization), all of which are known to
have effects on the synchronizability of oscillator networks. \ We find that
these topological factors produce marked signatures in the Laplacian
eigenvalue distribution and in the localization properties of individual
eigenvectors. \ Using a set of coordinates based on the Laplacian eigenvectors
as a diagnostic tool for synchronization dynamics, we find that the process of
frequency synchronization can be visualized as a series of quasi-independent
transitions involving different normal modes. \ Particular features of the
partially synchronized state can be understood in terms of the behavior of
particular modes or groups of modes. \ \ For example, there are important
partially synchronized states in which a set of low-lying modes remain
unlocked while those in the main spectral peak are locked.\ \ We find
therefore that spectra influence dynamics in ways that go beyond results
relating a single threshold to a single extremal eigenvalue. \ 

\end{abstract}
\pacs{89.75.Hc, 05.45.Xt}
\maketitle

\section{\bigskip Introduction}

In the recent intense activity that has focused on complex networks
\cite{Strogatz}-\cite{Watts}, \ one goal has been to identify properties of
networks that are important for their function and behavior. \ A number of
concepts have been introduced for classifying network structures, for example
degree distribution, path length\cite{SocialNets}, clustering\cite{WS}, the
small-world property\cite{WS}\cite{Watts}, modularity,
betweenness\cite{Betweenness}, etc. \ Diagnostic tools for assessing function
include percolation properties, robustness against node or link
deletion\cite{attackTol}, studies of epidemic spreading\cite{Epidemics}, and
various measures of synchronizabilty. \ Synchronization can be approached by
means of global order parameters\cite{Sync}\cite{Kuramoto} or stability
analyses of fully synchronized\cite{PecoraMSF} or fully
incoherent\cite{Kuramoto}\cite{Restrepo} states, which give thresholds for the
onset of either synchronization or desynchronization. \ 

One way of distilling information about a network is to analyze the eigenvalue
and eigenvector spectra of matrices associated with the network, and that is
the focus of this paper. \ Such spectral properties provide an important
intermediary between structural and dynamical properties. \ \ They are derived
directly from the network topology and many relations exist between particular
eigenvalues and important network structural properties\cite{Biggs}%
\cite{Mohar}\cite{Chungbook}, while a number of studies have related extremal
eigenvalues to dynamical properties and thresholds.\cite{PecoraMSF}%
\cite{Restrepo} In the current paper we use spectral tools to diagnose the
dynamics of partial synchronization, between the incoherent and fully coherent states.

The important matrices are derived from either the adjacency or coupling
matrix. \ \ The adjacency matrix $A_{ij}$ is defined by $A_{ij}=1$ if a
connection exists between the nodes numbered $i$ and $j$ and $A_{ij}=0$
otherwise. \ The coupling matrix $W_{ij}$ is simply the matrix of coupling
strengths among nodes, \ and if the links are all equally weighted then it is
a multiple of the adjacency matrix. \ Several definitions of the Laplacian
matrix are in use. We will define it here as%
\begin{equation}
L_{ij}=(\sum_{j}W_{ij})\delta_{ij}-W_{ij}. \label{Ldef1}%
\end{equation}
In the case where $W_{ij}=A_{ij}$, this matrix is sometimes known as the
"combinatorial Laplacian."\cite{ChungPNAS}\ \ The definitions and significance
of these matrices will be discussed further in the following sections.

Some applications of network matrices depend only on one or two extremal
eigenvalues. \ A widely known application of the Laplacian matrix is the
Master Stability Function (MSF) technique for analyzing the stability of \ a
synchronized state of coupled oscillators.\cite{PecoraMSF} \ For the MSF,
\ the quantity of interest is the ratio of the largest to the smallest nonzero
eigenvalue of the Laplacian, and accordingly many studies of the Laplacian
spectrum on complex networks are limited to tabulations of this ratio. \ In
mathematical graph theory, \ a number of theorems relate geometrical
properties such as the diameter of the network to the smallest nonzero
eigenvalue\cite{Mohar}. \ Likewise with the adjacency matrix, for some
applications it is only the largest eigenvalue that matters. \ \ For example,
under certain assumptions, approximate relations were derived between the
largest eigenvalue of the adjacency matrix and the critical coupling strength
for the onset of phase synchronization in a network of limit cycle
oscillators.\cite{Restrepo}

In many cases, however, there is additional important information contained in
the full spectrum and in the eigenvectors themselves. \ \ The Laplacian
spectrum, for example, is relevant to the solution of diffusion and flow
problems on networks \cite{Mohar}. \ Apart from the MSF formalism, \ it is
applicable more generally to the dynamics of coupled oscillators near the
synchronized state, \ including the relaxation of coupled identical
limit-cycle oscillators to equilibrium.\cite{Arenas} \ \ In a previous paper
\cite{McGrawSpec}, we showed that the eigenvectors of the Laplacian form a
useful coordinate system in which to view the dynamics of partly synchronized
networks of oscillators, \ even at a significant distance from full
synchronization. \ 

There have been relatively few efforts to study the full spectrum of the
Laplacian as defined in (\ref{Ldef1}) for general complex networks. \ In the
mathematical literature, more attention has often been paid to the adjacency
matrix\cite{Biggs} or the so-called "normalized Laplacian"\cite{Chungbook}%
\cite{ChungPNAS} which is related to $L$ but can have a quite different
spectrum. \ Studies of the combinatorial Laplacian\cite{Mohar} are not often
focused on applications to large complex networks. \ There have been some
numerical and analytical studies of the adjacency matrices of complex
networks\cite{Goh}\cite{Farkas}\cite{ChungPNAS}\cite{Dorogovtsev} as well as
some studies of the normalized Laplacian\cite{ChungPNAS} and of the closely
related "transition matrix"\cite{Dorogovtsev} on one category of random
uncorrelated networks with given expected degree distributions. As for the
Laplacian defined in (\ref{Ldef1}), \ its full spectrum has been examined on
random Erdos-Renyi \cite{Biroli} and small-world networks\cite{Monasson} but
much of the territory is still relatively uncharted, especially in the case of
networks with degree correlations, clustering, communities, and other types of
correlations. \ 

The current paper explores the information contained in the Laplacian
spectrum, and the three-way linkage among network topology, spectrum and
dynamics, in particular oscillator synchronization dynamics. \ \ One goal is
simply to characterize the Laplacian eigenvalues and eigenvectors for several
important types of networks. \ We \ pay attention to the spectral and
dynamical effects of network topological properties such as degree
distribution (especially Poisson vs. scale-free or homogeneous vs.
heterogeneous), clustering (i.e., the tendency of neighbors of a given node to
form links with each other), and community structure, also known as
modularity. \ While recent studies of network matrix spectra have dealt with
random network models without correlations, \ modularity and clustering
represent significant new ingredients. \ \ In addition, \ we consider
weighting and symmetry or asymmetry of connection strengths, especially the
asymmetric connection scheme in which the total input to each node is
normalized.\ We are interested in the latter coupling scheme because it is
known to optimize synchronizability in MSF terms. \cite{Normalization}%
\cite{Normalization2}

Beyond the mere classification of spectra, another goal is to understand
dynamics. \ For the types of networks studied, we consider the synchronization
dynamics of coupled phase oscillators (the Kuramoto\cite{Kuramoto} model),
using coordinates based on the Laplacian eigenbasis as a tool for visualizing
the dynamics and revealing structure within the synchronization transition.
\ This program follows the general inspiration of the MSF technique in that
the aim is to use the Laplacian spectrum to isolate topological influences on
synchronization, as distinct from the individual node dynamics. \ The MSF,
however, \ is strictly applicable only to the stability problem of a fully
synchronized state of a set of exactly identical oscillators. \ It is most
useful for networks of identical chaotic oscillators. \ For appropriately
limited questions, the MSF, relying only the eigenratio, gives rigorous
answers. \ The present work, on the other hand, \ seeks rather more heuristic
tools for much broader questions. \ \ We consider non-identical oscillators
and we consider the process of synchronization from its onset up to nearly
complete synchronization, \ rather than being limited to the immediate
neighborhood of the synchronization manifold. \ \ The current paper follows a
Brief Report\cite{McGrawSpec} in which a subset of our results was presented.
\ We expand on those results here by studying a greater variety and larger
sample of networks and by examining properties of the individual eigenvectors
such as localization and degree bias. \ \ 

The remainder of the paper is organized as follows: \ Section
\ref{Lproperties} discusses some general properties of the Laplacian matrix
that are relevant to network dynamics. \ Section \ref{Spectra} is devoted to a
description of particular network Laplacian eigenvalue spectra and their
dependence on topological properties. \ In section \ref{Shapes} we then
consider some properties of the eigenvectors themselves, especially
localization and degree bias. \ \ Finally, in section \ref{dynamics}, we
examine some connections between spectra and dynamics in the case of a network
version of the Kuramoto model. \ A coordinate system for phase space based on
the Laplacian eigenvectors proves useful for obtaining a geometric picture of
the oscillators' dynamical behavior. \ Under some conditions, \ groups of
eigenvectors behave as dynamically independent degrees of freedom and\ the
process of synchronization amounts to a contraction of phase space onto
progressively lower-dimensional submanifolds spanned by lower-eigenvalue
eigenvectors. \ The concluding section \ref{Conclusions} summarizes our
results and suggests some future directions. \ \ 

\section{Properties and Significance of the Laplacian\label{Lproperties}}

We begin by reviewing and extending some known general properties of the
Laplacian of a network of $N$ nodes. \ We denote the $N$-component
eigenvectors and associated eigenvalues respectively by $\mathbf{V}^{\alpha}$
and $\lambda_{\alpha}$, so that
\begin{equation}
\sum_{j=1}^{N}L_{ij}V_{j}^{\alpha}=\lambda_{\alpha}V_{i}^{\alpha}.
\label{Eigenvaluedef}%
\end{equation}
From the definition (\ref{Ldef1}) it follows that the sum of matrix elements
in any row is zero, and consequently the constant vector $(1,1,...1)$ is
always an eigenvector with eigenvalue zero. \ If the coupling matrix is
symmetric ($W_{ij}=W_{ji}$) then so is the Laplacian, and therefore all
eigenvalues are real and eigenvectors corresponding to different eigenvalues
are orthogonal. \ The trace of $L$ is given by
\begin{equation}
\mathrm{Tr}L=\sum_{i}L_{ii}=\sum_{j}\sum_{i}W_{ij}-\sum_{i}W_{ii}.
\label{Trace1}%
\end{equation}
(From now on, \ limits of summations are suppressed and assumed to be from 1
to $N$ unless otherwise indicated.) \ If self-couplings are excluded, then the
final term vanishes as $W_{ii}=0$. \ If, furthermore, \ $W_{ij}=A_{ij}$ (i.e.,
all couplings equally weighted) then the row sum of $W$ is just the degree (or
number of neighbors) of each node, and so
\begin{equation}
\sum_{\alpha}\lambda_{\alpha}=\mathrm{Tr}L=\sum_{i}k_{i}, \label{TraceID}%
\end{equation}
where $k_{i}=\sum_{j}W_{ij}$ is the degree of the $i$th node. \ Some authors
(e.g., \cite{Mohar}) \ define the quantity $\sum_{j}W_{ij}$ as the "degree"
even when the couplings $W_{ij}$ are not only zeros and ones. \ However, it is
useful to be able to distinguish the sum of couplings from the actual number
of nodes to which a given node is connected. \ In cases of possible confusion,
we suggest the term "topological degree" for the latter, and following the
nomenclature of Ref. \cite{Normalization2}, we use "intensity" to refer to the
sum of input coupling strengths. \ For general couplings, (\ref{TraceID})
remains valid if the degree is replaced by the intensity. \ This means that
the average of the Laplacian eigenvalues is equal to the average degree (or
intensity) $\left\langle k\right\rangle $ for the whole network, and
consequently $\left\langle k\right\rangle $ is a convenient scaling factor for
comparing spectra of different networks.\footnote{Another procedure is
necessary in cases where $<k>$ is not well-defined, such as the thermodynamic
limit of scale-free networks with small power-law exponents. \ \ } \ 

A useful identity is the following:%
\begin{equation}
\sum_{i,j}L_{ij}x_{i}x_{j}=\sum_{i,j}W_{ij}(x_{i}-x_{j})^{2},
\label{LapdiffID}%
\end{equation}
where $x_{i}$ may be any quantity associated with each node (in the context of
synchronization, it could be the phase of each oscillator). \ The above
guarantees that, as long as the couplings are non-negative, then the Laplacian
is positive semi-definite; i.e., \ all of its eigenvalues are positive or
zero. \ If $x_{i}$ represents a perturbation of some dynamical degree of
freedom (such as the phase of an oscillator) then (\ref{LapdiffID}) suggests a
heuristic interpretation of $L$ as a tensor expressing the rigidity of the
network against such perturbations. \ In particular, if the perturbation is a
multiple of one of the normalized eigenvectors, or in other words if%
\[
x_{i}=\eta V_{i}^{\alpha},
\]
then%
\begin{equation}
\sum_{i,j}W_{ij}(x_{i}-x_{j})^{2}=\lambda_{\alpha}\eta^{2}.
\label{LapdiffEigenvectorID}%
\end{equation}
The left-hand side of eq. (\ref{LapdiffEigenvectorID}) is a sum of squared
differences across links, weighted by the strength of each link; it can be
imagined as a sum of potential energies due to stretched bonds. If the
eigenvalue $\lambda_{\alpha}$ is small, then the perturbation along the
corresponding vector can occur without disturbing very many strong bonds, and
the opposite is true if $\lambda_{\alpha}$ is large. \ With this
interpretation in mind, the zero mode $(1,1,...1)$ indicates the freedom of a
uniform translation of the whole network (in the case of coupled oscillators,
it is the freedom to reset all phases by an equal amount without affecting any
links.) \ 

The definition (\ref{Ldef1}) \ shows a deceptively simple relationship between
the adjacency (or coupling) matrix and the Laplacian. \ However, it is only in
the special case of a so-called \emph{regular network} (i.e., one in which the
degree or the sum $K=\sum_{j}W_{ij}$ is the same for all $i$), \ \ that the
two matrices commute and thus are guaranteed to share a common eigenbasis.
\ \ In this special case, \ the corresponding eigenvalues $\lambda_{\alpha}$
for the Laplacian and $\mu_{\alpha}$ for the coupling matrix are related by
$\lambda_{\alpha}=\mu_{\alpha}-K$. \ \ \ \ In general, \ it is easy to show
that the commutator%
\begin{equation}
\left[  \mathbf{W,L}\right]  \equiv\mathbf{WL-LW} \label{Commutator}%
\end{equation}
is a matrix whose elements are proportional to the differences between the
degrees of adjacent nodes. \ \ This leads us to expect that the differences in
spectrum between the two matrices become more important as the network becomes
more heterogeneous, \ \ but on the other hand can be lessened by what is
called assortative mixing\cite{Assortative} (i.e., a tendency of nodes to
connect with other nodes of similar degree.) \ 

\section{Topology and Laplacian Spectra\label{Spectra}}

\begin{figure}
[ptb]
\begin{center}
\includegraphics[
height=3.7576in,
width=3.0381in
]%
{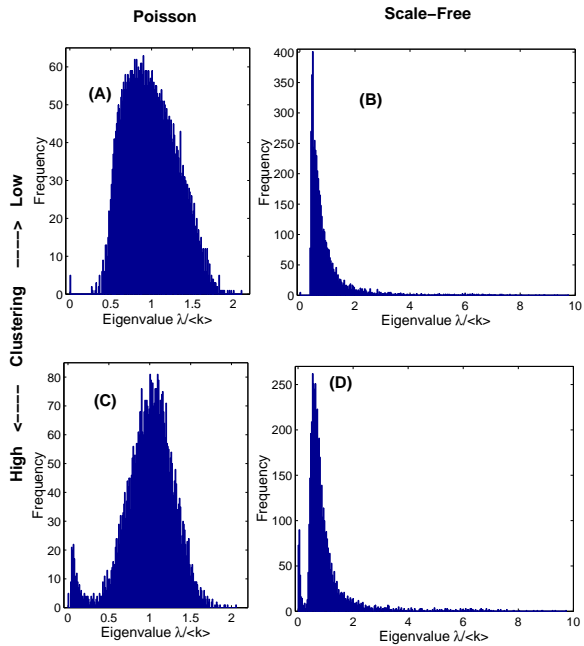}%
\caption{Histograms of scaled Laplacian eigenvalues for Poisson and scale-free
networks with low and high clustering coefficients. \ Each histogram is
cumulative for five different networks drawn from the same ensemble. \ }%
\label{combhist}%
\end{center}
\end{figure}

In this section we examine the shapes of the Laplacian spectra of several
types of networks, and consider how specific features of the spectra are
correlated with structural features of the networks. We will begin with
networks in which all existing links are bidirectional and equally weighted,
and then consider certain types of asymmetries. \ Except where otherwise
indicated, the networks we study have $N=1000$ nodes and average topological
degree $\langle k\rangle=20$. \ Figures \ref{combhist} A,B show histograms of
the (scaled) Laplacian eigenvalue distributions for random (Erdos-Renyi
\cite{ERmodel}) networks with Poisson degree distribution (henceforth referred
to as a Poisson network or PN) and Barabasi-Albert\cite{BANetworks} networks
with scale-free distribution (a scale-free network or SFN). \ The first thing
to note is the dependence of the spectrum on the degree distribution. \ The
eigenvalue distributions share some statistical features of the degree
distributions: for example, the eigenvalue spectrum of the SFN has a power-law
tail just as does the degree distribution. \ For the Poisson networks, on the
other hand, the spectrum has a Poisson-like single peak with no significant
tail. \ When we varied $\left\langle k\right\rangle $ from 5 to 40 (plots are
not shown here), we found that the relative width (or width scaled by
$\left\langle k\right\rangle $) of this peak decreases with increasing
$\left\langle k\right\rangle $ but that other qualitative features are the
same. \ \ \

We next examine the effect of clustering, which is known to have strong
effects on the synchronizability of networks.\cite{McGraw} \ We applied a
stochastic rewiring algorithm \cite{Kim}\cite{McGraw}to increase the
clustering coefficient of the PN and SFN while leaving the degree
distributions unchanged. \ Clustering (also called transitivity) refers to the
tendency of two nodes which share a common neighbor to have an increased
likelihood of also being directly connected to each other (compared to two
nodes that do not share a neighbor)\cite{WS}. \ \ Put another way, clustering
indicates the prevalence of triangles in the network topology. \ The
clustering coefficient $\gamma$, a numerical measure of clustering, is defined
as an average over the network of the local clustering coefficient, given by
\begin{equation}
\gamma_{i}=\frac{t_{i}}{\binom{k_{i}}{2}}=\frac{2t_{i}}{k_{i}(k_{i}-1)},
\label{Clusterdef}%
\end{equation}
where $t_{i}$ is the number of mutual connections among the neighbors of a
given node, \ $k_{i}$ is the number of neighbors, and $\binom{k_{i}}{2}$\ is
the number of possible pairs of neighbors that could potentially be connected.
\ The bottom row of plots in figure \ref{combhist} show the spectra of
networks with the same degree distributions as the ones above, but with high
clustering coefficients, specifically $\gamma=0.640\pm0.005$ for the PN's and
$\gamma=0.675\pm0.005$ for the SFN's. \ For the "natural" low-clustering
networks whose spectra are in the upper row of figure \ref{combhist}, \ the
values are within the range $\gamma=0.0195\pm0.0005$ for the PN's and
$\gamma=0.073\pm0.005$ for the SFN's. \ \ It is apparent that for both the PN
and SFN, increasing the clustering changes the shape of main spectrum
slightly, sharpening the peak and shifting it to the right, \ while also
creating a new group of eigenvalues close to zero. \ A series of plots at
intermediate values of $\gamma$ (not given here) shows that as $\gamma$
increases, this group of eigenvalues breaks away from the main peak and
gradually migrates downward toward zero. \ 

\begin{figure}
[ptb]
\begin{center}
\includegraphics[
height=4.6985in,
width=3.039in
]%
{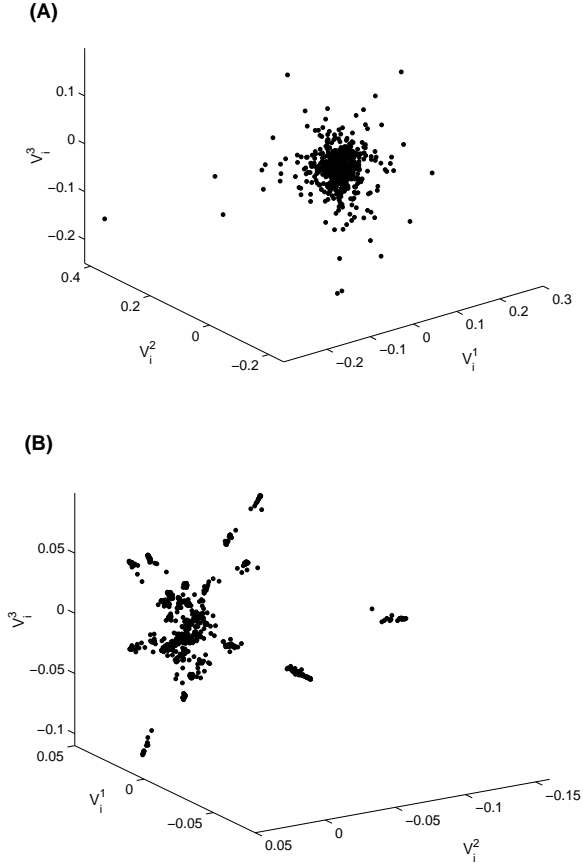}%
\caption{Scatter plots of the components of the Laplacian eigenvectors with
the lowest three nonzero eigenvalues for scale-free networks with low (top)
and high (bottom) clustering. \ In the high-clustering case, communities
appear as clumps of nodes. \ }%
\label{donettiplots}%
\end{center}
\end{figure}

The presence of near-zero eigenvalues generally indicates the existence of
strong communities, \ or nearly disconnected components. \ The multiplicity of
the (exactly) zero eigenvalue is equal to the number of disconnected
components \cite{Mohar}. Strong communities (subsets of nodes with much fewer
connections between groups than within groups) \ behave like nearly
disconnected components and thus result in small but nonzero eigenvalues.
\ Low eigenvalues and the corresponding eigenvectors have been used in some
algorithms for partitioning and/or detection of communities. \ The most common
application is to find an optimal partitioning of the network into two
communities by means of the eigenvector corresponding to the second-lowest (or
lowest non-zero) eigenvalue, a procedure which may be applied
iteratively.\cite{Spectral Comm} \ \ More recently, however, \ techniques for
finding communities have made use of the space spanned by the lowest few
eigenvectors rather than only one.\cite{Donetti}\cite{CapocciComm} \ The
essential technique is as follows: \ the $m$ eigenvectors with the lowest
nonzero eigenvalues are found, \ and each node is assigned $m$ coordinates
which are the entries of the $m$ eigenvectors at the position of that node.
\ \ When these coordinates are plotted for all nodes, \ communities, if they
exist, \ appear as groups of nodes clumped together in this $m$-dimensional
space. \ The clumps may be readily observable by eye; \ alternatively, one may
use statistical techniques such as hierarchical clustering\cite{Donetti} to
find the optimal grouping into clumps. \ We have made such a plot for the
first three eigenvectors of the high- and low-clustering SFN's in Fig.
\ref{donettiplots}. \ \ The plot clearly shows that the nodes of the
high-clustering network group together into communities while those of the
low-clustering network do not. \ In examining synchronization dynamics in
section \ref{dynamics}, the low-lying eigenvectors will give a good indication
of the dynamical importance of communities. \ In view of the interpretation of
$L$ as giving the network's inherent rigidity against perturbations, \ the
low-lying eigenvectors associated with communities reflect the relative ease
with which the network can fall apart (or desynchronize) along community
boundaries. \

Although in network studies it it most common to consider networks with
symmetric, undirected links, \ Motter et al. \cite{Normalization}%
\cite{Normalization2} have shown for a variety of topologies that the MSF
synchronizability as measured by the eigenratio $\lambda_{N}/\lambda_{2}$ is
improved by a particular scheme of asymmetric coupling which we refer to here
as input normalization or simply \emph{normalization}. \ This is a weighting
in which the input coupling strengths to any node are scaled so that their sum
is unity. \ \ Put another way, \ the coupling matrix is derived by dividing
each row of the adjacency matrix by the degree of the corresponding node:%
\begin{equation}
W_{ij}=A_{ij}/k_{i}=\frac{A_{ij}}{\sum_{j}A_{ij}}. \label{normalization}%
\end{equation}
In a heterogeneous network, this implies that the couplings are not symmetric.
\ If a hub node H with high degree is connected to a peripheral node P with
low degree, then the influence of H on P is greater in absolute terms than
that of P on H. \ \ P's influence on H is diluted by H's many other neighbors.
\ \ Motivated by the enhanced synchronizability of normalized networks, \ we
are interested in understanding the effect of normalization on the full
Laplacian spectrum (and subsequently on dynamics), \ not only the extremal
eigenvalues. \ \ In spite of the asymmetry introduced into the coupling and
therefore the Laplacian, it can be proved \cite{Normalization} that the
Laplacian eigenvalues are still all real in this case. \ Furthermore, the
intensities of all nodes are unity, \ so that, as discussed above, \ the
Laplacian has a common eigenbasis with the coupling matrix. \ \ (Due to the
asymmetry, however, the eigenvectors are not mutually orthogonal.) \ The
eigenvalue spectrum is in fact the same as for what is known in graph theory
as the "normalized Laplacian," \cite{Chungbook}\cite{ChungPNAS} of the
underlying unweighted network, although the eigenvectors themselves are not
the same. \ To summarize the relationships among the matrices, let
$\mathbf{D}$ be the diagonal matrix whose entries are the topological degrees,
let $\mathbf{L}_{u}$ and $\mathbf{L}_{n}$ be the Laplacians of the
unnormalized and normalized networks repectively, while $\mathcal{L}$ is the
symmetric "normalized Laplacian." Then the relationships are: \
\begin{align}
\mathbf{L}_{u}  &  =\mathbf{D-A}\\
\mathbf{L}_{n}  &  =\mathbf{L}_{u}\mathbf{D}^{-1}\nonumber\\
\mathcal{L}  &  =\mathbf{D}^{-1/2}\mathbf{L}_{u}\mathbf{D}^{-1/2}.\nonumber
\end{align}

\begin{figure}
[ptb]
\begin{center}
\includegraphics[
height=3.4964in,
width=3.039in
]%
{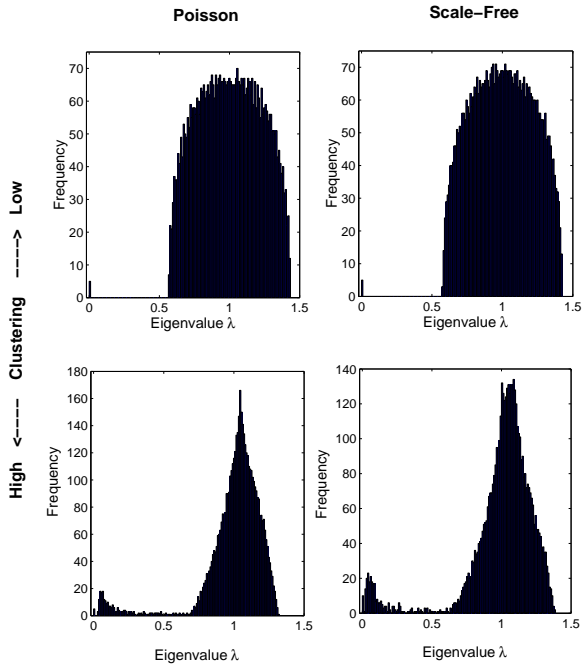}%
\caption{Eigenvalue histograms for networks with normalized inputs (Eq.
\ref{normalization}). \ The networks are topologically the same as in figure
\ref{combhist}; \ only the connections strengths have been changed. \ \ Since
all nodes have a coupling sum of unity, the scaling factor $\langle k\rangle$
is not needed here: \ the eigenvalues $\lambda$ in figs. \ref{combhistn}%
,\ref{iprncomb.eps} and \ref{devncomb} are comparable to the scaled values
$\lambda/\langle k\rangle$ in figs. \ref{combhist}, \ref{iprcomb} and
\ref{devcomb}.}%
\label{combhistn}%
\end{center}
\end{figure}

Figure \ref{combhistn} shows eigenvalue histograms for the same networks as in
Figure \ref{combhist}, but with the inputs normalized. \ The first striking
feature is that the spectra look very nearly the same for both the PN and the
SFN. \ \ As was found in \cite{Normalization2}, the effect of degree
heterogeneity is greatly suppressed by normalization; \ we will see that this
is also true of the effect of degree distribution on dynamics. \ In the low
clustering case, the spectra have approximately semicircular shapes for both
networks, \ recalling the so-called semicircle law\cite{Wignerlaw} for spectra
of random matrices. \ \ Previously, the normalized Laplacians of a broad class
of uncorrelated random matrices (including the Erdos-Renyi but not, strictly
speaking, the Barabasi-Albert network) were found to obey the semicircle
law\cite{ChungPNAS}, as were the closely related transition
matrices\cite{Dorogovtsev}. \ \ Consistent with what is known about the
eigenratio\cite{Normalization}, \ the spectra are much narrower than for the
unnormalized networks, \ with no apparent tail either of the exponential or
power-law type. \ \ 

Increasing the clustering of the normalized networks narrows and markedly
sharpens the main spectral peak into a more triangular shape. \ \ The other
main effect of clustering, namely the creation of a second peak near zero,
\ is much the same as in the unnormalized networks. \ This is consistent with
our understanding of the low modes as being associated with community
structure. \ Communities are a topological phenomenon, \ the result of a
paucity of connections among the different subsets. \ \ If there are only a
few connections between two subsets of the network, then an adjustment of
coupling \emph{strengths }alone is unlikely to compensate significantly for
this scarcity, unless some bias results in extra strengthening of
inter-community ties at the expense of others. \ Input normalization certainly
alters the dynamical roles of high- vs. low-degree nodes, but it is not
surprising that it has little effect on such global topological features as
community formation. \

\section{Shapes of the Eigenvectors: localization and
delocalization\label{Shapes}}

\begin{figure}
[ptb]
\begin{center}
\includegraphics[
height=4.4581in,
width=3.1306in
]%
{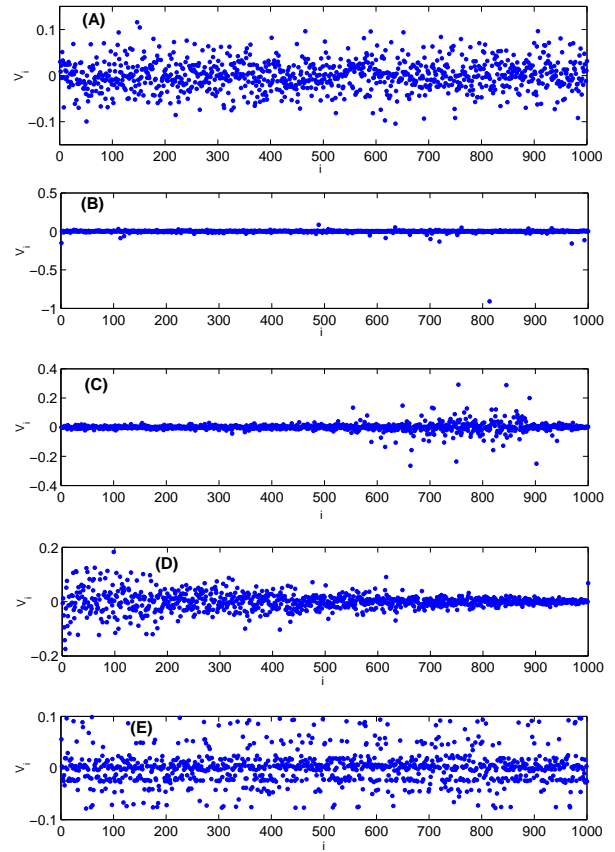}%
\caption{Examples of different shapes of Laplacian eigenvectors. \ Each is
typical for one type of network and a particular range of eigenvalues. \ \ $x$
coordinates are the node index, \ where the nodes have been sorted from lowest
to highest degree; \ $y$ coordinates are the eigenvector components. \ (A)
Low-clustering PN, vector $\alpha=500$ (i.e., the vector with the 500th lowest
eigenvalue), $\lambda=19.27$. This vector is delocalized and its components
appear random. \ \ (B) Low-clustering PN, $\alpha=4$, $\lambda=7.97$. This
vector is strongly localized: note the single large component at $i\approx810$
\ (C) Low-clustering SFN, $\alpha=800$, $\lambda=23.74$. Not strongly
localized, but most large components occur at nodes within a medium to high
range of degrees.\ (D) Low-clustering SFN, $\alpha=50$, $\lambda=8.26$.
\ Largest components occur at low-degree nodes. \ (E) \ High-clustering PN,
$\alpha=4$, $\lambda=0.86$. \ This is an example of the low-lying modes that
form in networks with high clustering. \ Components fall near a small number
of discrete values, giving the plot a striated appearance. }%
\label{vectorexamples}%
\end{center}
\end{figure}

To the extent that different Laplacian eigenvectors are associated with
collective degrees of freedom having different dynamical functions, \ it is
interesting to examine the "shapes" of these vectors and understand how they
relate to the network's structure. \ In figure \ref{vectorexamples}, we plot
the components $V_{i}^{\alpha}$ of several representative eigenvectors against
the node index $i$ where the nodes have been sorted from lowest to highest
degree. \ While all plots appear partly random, reflecting the randomness in
the network's structure, \ there are nonetheless patterns that are clear from
a visual inspection, and clear qualitative differences between different
vectors. \ Some vectors, such as the one shown in Fig. \ref{vectorexamples}B,
\ are highly localized, with only a few large nonzero components and the rest
nearly zero, while others such as the one in frame A have nonzero components
spread throughout the whole network. \ \ Some, such as C or D, \ show a degree
bias; \ most of their large components occur at nodes within a certain degree
range (either high, low or intermediate). \ \ The vector in E is an example of
one of the low-lying modes in a strongly clustered network. \ The components
cluster around a small number of discrete values. \ Nodes having the same
value of $V_{i}$ are likely to belong to the same community. \ (As described
above, the communities are separated more distinctly if several low-lying
eigenvectors are plotted simultaneously.) \ \

A convenient scalar measure of a vector's degree of localization is the
so-called inverse participation ratio\cite{IPRDef} (IPR) $P$, which is defined
for any vector $\mathbf{V}$ by
\begin{equation}
P(\mathbf{V})=\frac{\sum_{i}V_{i}^{4}}{(\sum_{i}V_{i}^{2})^{2}}.
\label{IPRDef}%
\end{equation}
(Note that the denominator is 1 if the vector is normalized.) \ \ $P$ ranges
from a minimum value of $1/N$ (for a normalized vector whose components are
all of equal magnitude $1/\sqrt{N}$) to a maximum of $1$ for a vector with
only one nonzero component. \ The more localized the vector (ie., the less
evenly its weight is spread among multiple components) \ the higher the value
of $P.$ \ In order to quantify the degree bias noted for some of the
eigenvectors, let us define the degree expectation value (DEV) $Q$ for a
vector as%
\begin{equation}
Q(\mathbf{V})=\frac{\sum_{i}V_{i}^{2}k_{i}}{\sum_{i}V_{i}^{2}}. \label{DEVdef}%
\end{equation}
(Some other authors\cite{Biroli} have used instead the "center connectivity,"
which is simply the degree of the node with the maximum value of $\left\vert
V_{i}\right\vert $. \ The center connectivity should be nearly identical to
the DEV in cases of vectors localized very strongly at a single node, but $Q$,
being an average, is likely to be a more robust and meaningful metric in cases
such as panel (C) of Fig. \ref{vectorexamples} where a vector has some level
of localization and degree bias but no single component is dominant.) \ 

\begin{figure}
[ptb]
\begin{center}
\includegraphics[
height=4.0222in,
width=3.2897in
]%
{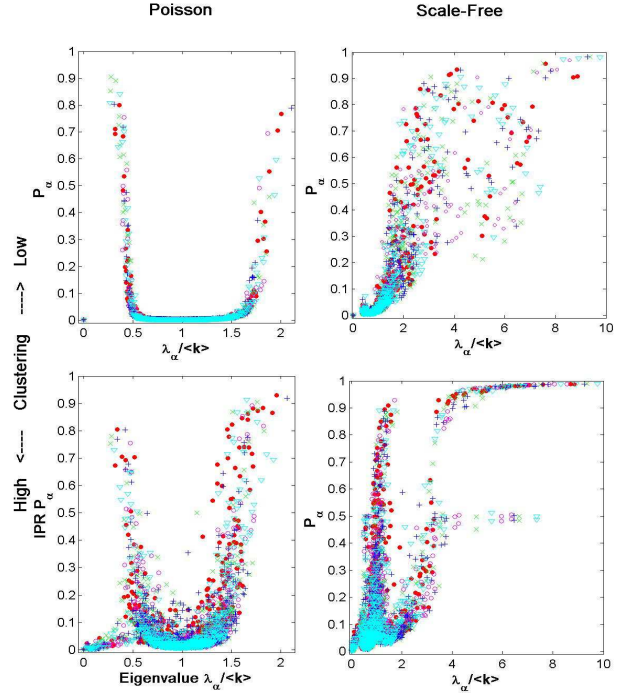}%
\caption{(color online) Inverse participation ratios $P_{\alpha}$ (eq.
\ref{IPRDef}) of Laplacian eigenvectors for unnormalized networks, plotted
against the corresponding scaled eigenvalues. \ Here and in all of figs.
\ref{iprcomb}-\ref{devncomb}, different symbols (colors) are for five
different networks from the same ensemble. \ }%
\label{iprcomb}%
\end{center}
\end{figure}

\begin{figure}
[ptbptb]
\begin{center}
\includegraphics[
height=3.9963in,
width=3.2889in
]%
{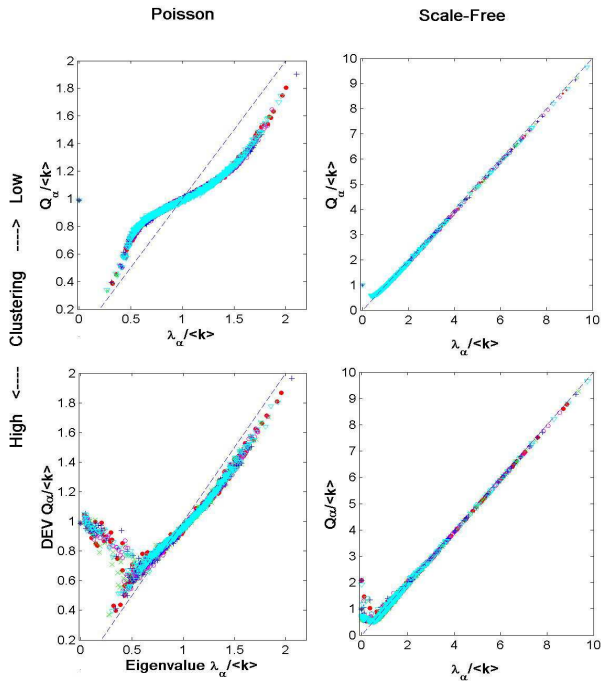}%
\caption{(color online) \ Scaled degree expectation values $Q_{\alpha}/\langle
k\rangle$ (eq. \ref{DEVdef}) for Laplacian eigenvectors of unnormalized
networks. \ Eigenvalues and DEV's are strongly correlated in all cases (a
dotted line with slope 1 is shown for comparison). \ }%
\label{devcomb}%
\end{center}
\end{figure}

\begin{figure}
[ptbptbptb]
\begin{center}
\includegraphics[
height=4.0136in,
width=3.2897in
]%
{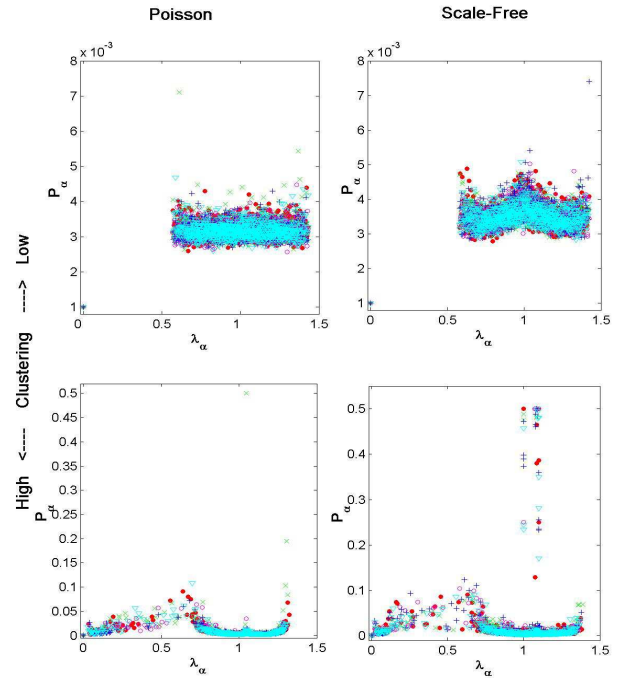}%
\caption{(color online) Inverse participation ratios $P_{\alpha}$ of Laplacian
eigenvectors for normalized networks.}%
\label{iprncomb.eps}%
\end{center}
\end{figure}

\begin{figure}
[ptbptbptbptb]
\begin{center}
\includegraphics[
height=3.7913in,
width=3.2897in
]%
{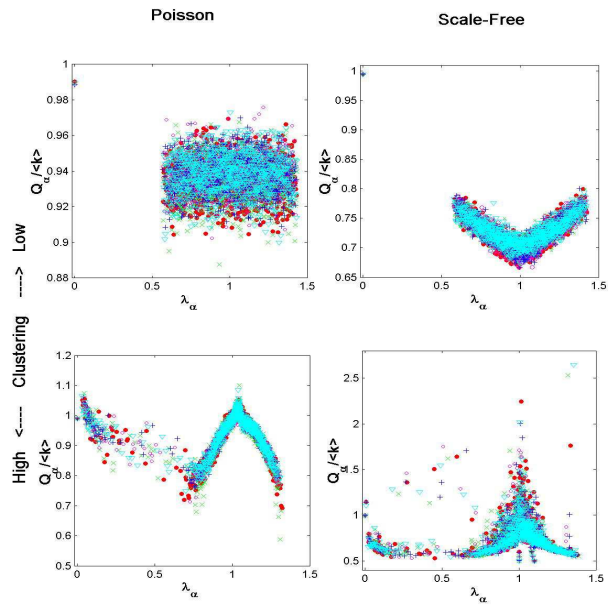}%
\caption{(color online) Degree expectation values $Q_{\alpha}$ for Laplacian
eigenvectors of normalized networks.}%
\label{devncomb}%
\end{center}
\end{figure}

Figures \ref{iprcomb}-\ref{devncomb} show values of $P$ and $Q/\langle
k\rangle$ (i,e., the DEV scaled by the average degree) for the Laplacian
eigenvectors of several networks, plotted against the corresponding scaled
eigenvalues $\lambda/\langle K\rangle$. \ (Note that while $Q$ is
appropriately scaled by the average topological degree, the eigenvalues are
scaled by the average intensity, which is equal to the topological degree for
the unnormalized networks but unity for the normalized ones.) \ In the plots
for the unnormalized networks (figs. \ref{iprcomb} and \ref{devcomb}), several
features are notable. \ Among modes within the main spectral peak, \ those
near the edges tend to be more strongly localized. \  For both Poisson and
scale-free networks, increased clustering tends to increase the localization
of modes, especially in the tails of the eigenvalue distributions. \ The DEV's
are positively correlated with the eigenvalues, \ especially in the scale-free
case and for the tails of the eigenvalue distributions. \ An exception to
these trends is the group of low-lying modes that form at high clustering.
\ These modes are generally rather delocalized and have $Q$ close to the
average degree of the network. \ \ This is consistent with their
interpretation as reflecting global community divisions of the network,
involving most of the nodes. \ We note that our results for the spectra, IPR's
and DEV's for the low clustering Poisson networks agree with random network
results reported elsewhere\cite{Biroli}. \ The other results are largely
consistent with the general principle that localization tends to occur near
"defects" or nodes with degrees significantly higher or lower than
average.\cite{Biroli} \ A notable exception occurs for the strongly clustered
SFN, where there is a large group of  moderately to strongly localized modes
with scaled eigenvalues near unity (see fig. \ref{iprcomb}). \ From fig.
\ref{devcomb} it is evident that the DEV's of these modes fall near the
average of the degree distribution, not the tails. \ 

Analogous plots of eigenvector properties for normalized networks (figs.
\ref{iprncomb.eps} and \ref{devncomb}) show radical differences from the
unnormalized case. \ First, there is much less localization of modes, even at
high clustering. \ \ The dependence of DEV on eigenvalue is much weaker and
not monotonic. \ \ Rather, the main peak of the spectrum is approximately
symmetric about $\lambda=1$. \ In the normalized network unlike the
unnormalized case, the Laplacian spectrum is directly related to that of the
coupling matrix, and a Laplacian eigenvalue of 1 corresponds to a coupling
matrix eigenvalue of zero. \ It is a general property of random graphs that
their adjacency matrix spectra are approximately symmetric about zero
(deviations from symmetry occur when there are correlations or
clustering)\cite{Goh}. \ The spectra of normalized networks evidently share
this approximate symmetry, although the coupling matrix in this case is
\emph{not} the same as the adjacency matrix. \ It should be kept in mind that
the eigenvectors of $L$ for a normalized network are not orthonormal. \ 

In contrast to the main peak, the low-lying modes of strongly clustered
networks behave qualitatively much like those in unnormalized networks,
suggesting again that community structure is scarcely affected by
normalization. \ \ \ \ \

\begin{figure}
[ptb]
\begin{center}
\includegraphics[
height=2.8228in,
width=3.1401in
]%
{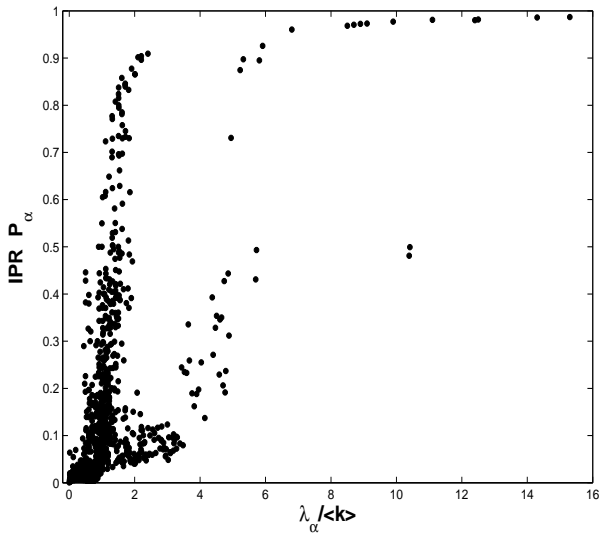}%
\caption{Inverse participation ratios vs. eigenvalues for the Laplacian
eigenvectors of a highly clustered SFN with $N=1500$, $\langle k\rangle=10$
and clustering coefficient $\gamma=0.560$. \ The plot is quite similar to that
in figure \ref{iprcomb} despite the different network sizes and
connectivities. \ }%
\label{smallkipr}%
\end{center}
\end{figure}

Without making a thorough analysis of the dependence of spectra on $\langle
k\rangle$ or the finite size $N$, we note that important qualitative features
appear to be independent of $\langle k\rangle$. \ As an example of this
observation, \ in figure \ref{smallkipr} we plot inverse participation ratios
versus eigenvalues for the Laplacian eigenvectors of a strongly clustered SFN
with $N=1500$, $\langle k\rangle=10$ and clustering coefficient $\gamma
=0.560$. \ \ This may be compared with the corresponding plot in figure
\ref{iprcomb} for the case $\langle k\rangle=20$; \ the peculiar shapes match
quite closely. \

\section{Spectra, Eigencoordinates and dynamics\label{dynamics}}

In this section, \ we examine numerically the synchronization behavior of a
network Kuramoto\cite{Kuramoto} model on the networks we have been studying.
\ Topological features (degree distribution, clustering, normalization) have
strong and sometimes complex effects on synchronization.\cite{McGraw}%
\cite{Normalization}\cite{Normalization2}\cite{PecoraMSF} We show that
projections onto the Laplacian eigenbasis are useful in understanding these
effects, and that specific dynamical behaviors of the networks are associated
with specific sets of modes in the Laplacian spectrum. \ This is true even in
strongly nonlinear regimes of partial synchronization, despite the fact that
the Laplacian is most naturally applied to linear problems near full synchronization.

We first define the model and show how the Laplacian and its spectrum appear
naturally in a linearized description of the frequency-synchronized state,
\ and then we proceed to use the Laplacian eigenvectors to parametrize the
partially desynchronized states, showing that this coordinate system remains
useful well beyond the range of validity of the linearization and that
individual modes may behave as quasi-independent degreess of freedom. \ 

Our model\cite{Moreno} is defined by the coupled equations%
\begin{equation}
\frac{d\phi_{i}}{dt}=\omega_{i}+\frac{\beta}{\left\langle K\right\rangle }%
{\displaystyle\sum\limits_{j}}
W_{ij}\sin(\phi_{i}-\phi_{j}), \label{Kuramotomodel}%
\end{equation}
where $\phi_{i}$ are $N$ phase variables (one associated with each node of a
network), \ $-1\leq\omega_{i}\leq1$ are the randomly and uniformly distributed
intrinsic frequencies, $\beta$ is the overall coupling strength, and $W_{ij}$
as before is the weighting matrix of the individual couplings. \ The coupling
strength is scaled by the average intensity $\left\langle K\right\rangle $ of
all nodes. \ In our simulations, we imposed the condition $\overline{\omega
}=0$ by subtracting the average from each realization of the random
frequencies. \ (This condition can be imposed without loss of generality; it
amounts to a transformation to a rotating frame of reference.) \ If the
couplings are symmetric, then the velocities $\frac{d\phi_{i}}{dt}$ obey the
exact sum rule
\begin{equation}
\sum_{i}\frac{d\phi_{i}}{dt}=\sum_{i}\omega_{i}=0 \label{SumRule}%
\end{equation}
due to the antisymmetry of the sine coupling function: \ when summed over $i$,
the coupling terms in equation (\ref{Kuramotomodel}) cancel. \ Note that eq.
(\ref{SumRule}) does not necessarily hold for a normalized network, due to the
asymmetry of the coupling $W_{ij}$. \ 

If the system is strongly synchronized so that all phase differences are
small, then the sine function can be linearized and the equations of motion
become approximately \ \
\begin{equation}
\frac{d\phi_{i}}{dt}=\omega_{i}+\frac{\beta}{\left\langle K\right\rangle }%
{\displaystyle\sum\limits_{j}}
W_{ij}(\phi_{i}-\phi_{j})=\omega_{i}-\frac{\beta}{\left\langle K\right\rangle
}%
{\displaystyle\sum\limits_{j}}
\pounds _{ij}\phi_{j}. \label{Linear EOM}%
\end{equation}
Thus the Laplacian matrix appears naturally in the description of small
deviations from full synchronization. \ 

\begin{figure}
[ptb]
\begin{center}
\includegraphics[
height=3.6547in,
width=3.4402in
]%
{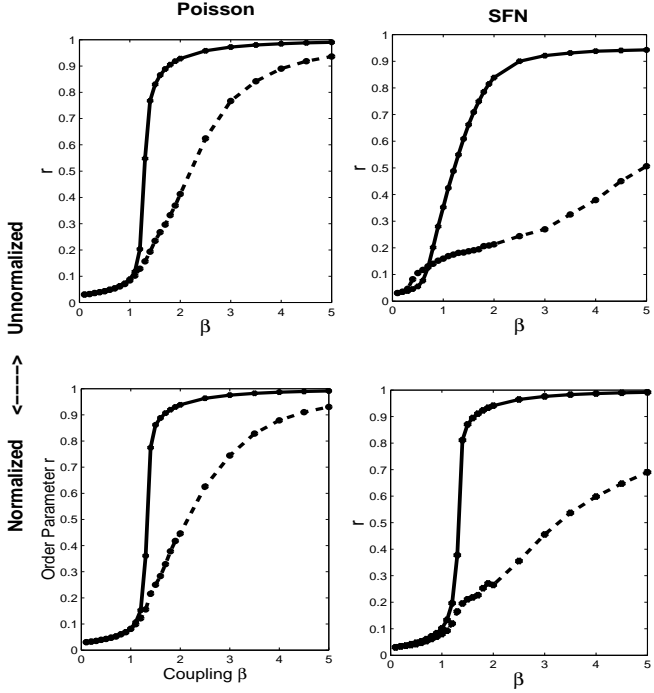}%
\caption{Global order parameter vs. coupling strength for Poisson and
scale-free networks with low and high clustering. Solid line: low clustering;
\ dashed line: high clustering. \ For the unweighted SFN's, note that there is
a range of low couplings $0.3\lesssim\beta\lesssim0.7$ for which the highly
clustered network is more synchronized than the low-clustering one. \ }%
\label{orderpar}%
\end{center}
\end{figure}

We simulated the model (\ref{Kuramotomodel}) using a fourth-order Runge-Kutta
method with time step 0.1. \ \ In figure \ref{orderpar} we show the global
synchronization order parameter
\begin{equation}
r=\left\langle \left\vert \sum_{j}e^{i\phi_{j}}\right\vert \right\rangle _{T}
\label{orderparam}%
\end{equation}
(Where $\left\langle {}\right\rangle _{T}$ signifies a time average) \ as a
function of the coupling strength $\beta$ for all of the network types
considered in the previous sections. \ In the simulations, the time average
was approximated by sampling 30 times at intervals of 10 time units, after
first runnimg for 200 time units in order to reach a steady state. \ The
results were then averaged over 10 realizations of the random frequency
distribution for each of five networks drawn from the ensemble. \ \ Note that
in all cases, increasing the clustering strongly suppresses full
synchronization, as shown by the fact that the order parameter curves for the
highly clustered networks approach unity much more slowly at large coupling
strength. \ This effect is more pronounced for the SFN than for the PN,
\ whether normalized or not. \ For the unnormalized SFN, \ however,
\ increased clustering has the seemingly contradictory effect of promoting the
onset of partial synchronization even though it inhibits full synchronization.
\ This effect, noted previously in \cite{McGraw} \ and confirmed by finite
size scaling analysis\cite{GomezMorenoIJBC} is apparent in the early upward
turn in the order parameter curve at $\beta\approx0.5.$ \ It was found
previously that it is the highest-degree nodes which synchronize first.
\ \ This advanced partial synchronization disappears, however, in the case of
normalized inputs. \ Since the advanced synchronization is associated with a
special behavior of the high-degree nodes, \ its disappearance is consistent
with the general observation that input normalization greatly reduces the
effects of degree heterogeneity. \ \

In \cite{McGraw} it was shown that nodes in different parts of the degree
distribution can play different dynamical roles: for example, \ in the case of
the SFN it is the hubs (high-degree nodes) that synchronize first. \ A
complementary way of viewing the synchronization dynamics is to examine it in
terms of appropriately chosen collective degrees of freedom rather than the
behavior of individual oscillators. \ Here as in \cite{McGrawSpec}, we define
collective coordinates by means of projections onto eigenvectors of the
Laplacian. \ \ We define projections of the phase and frequency vectors onto
these eigenvectors by
\begin{equation}
\phi^{\alpha}\equiv\sum_{i}\phi_{i}V_{i}^{\alpha},\;\omega^{\alpha}\equiv
\sum_{i}\omega_{i}V_{i}^{\alpha},\label{NormalCoord}%
\end{equation}
\ The \emph{normal coordinates} \ $\phi^{\alpha}$ are the appropriate ones for
describing the relaxation to equilibrium of a strongly synchronized system
\cite{McGraw}\cite{Arenas}. In addition, we define the observed frequencies
(rotation numbers) of the oscillators as the time averages%
\begin{equation}
\Omega_{j}=\left\langle \frac{d\varphi_{j}}{dt}\right\rangle _{T}%
.\label{Obsfreqdef}%
\end{equation}
Projecting the vector of observed frequencies onto the Laplacian eigenbasis
gives a time-averaged velocity along the direction defined by each
eigenvector:%
\begin{equation}
\Omega^{\alpha}=\sum_{j}\Omega_{j}V_{j}^{\alpha}.\label{velocitydef}%
\end{equation}
For the purpose of elucidating the network dynamics, \ we found these
\emph{normal velocities} \ more useful to work with than the normal
coordinates themselves for two reasons. \ First, \ from a computational point
of view, \ it is easier to construct time averages of the velocities as one
requires only initial and final phases (not modulo 2$\pi$ for this purpose)
together with the elapsed time. \ Second, we found that plots of the
velocities as functions of coupling strength displayed sharper transitions and
clearer patterns than averages (or mean squares) of coordinates. \
\begin{figure}
[ptb]
\begin{center}
\includegraphics[
height=2.9075in,
width=2.936in
]%
{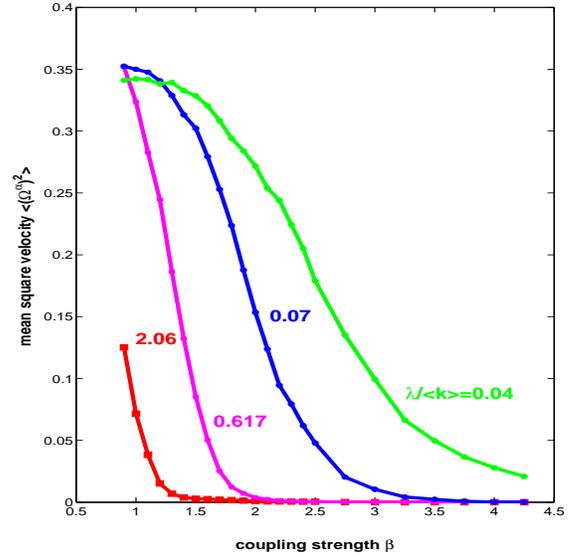}%
\caption{(color online): \ \ Plots of selected mean square normal velocities
\ $\langle(\Omega^{\alpha})^{2}\rangle_{\omega}$ as functions of coupling
strength $\beta$ averaged over 300 time units and 100 realizations of the
intrinsic frequencies $\omega$, for a single highly clustered Poisson network.
\ \ Each curve is labelled with the scaled eigenvalue corresponding to
$\alpha$. \ \ The two lowest (and farthest left) curves correspond to
eigenvalues at the high and low ends of the main spectral peak, while the
other two curves are for modes within the low-lying second peak of "community"
modes. \ \ The different velocity components drop to zero at different rates,
\ and ones corresponding to high eigenvalues lock (or drop to zero) at lower
coupling strengths.}%
\label{partialdisorder}%
\end{center}
\end{figure}

Figure\ \ref{partialdisorder} shows plots of the mean square velocities
$\langle(\Omega^{\alpha})^{2}\rangle$ as functions of coupling strength
$\beta$ for four different values of $\alpha$ taken from different parts of
the spectrum for a highly clsutered Poisson network. \ The time averages
$\Omega^{\alpha}$ were taken over an interval of 300 time units and the
squares were averaged over 100 realizations of the random frequencies. \ The
different $\langle(\Omega^{\alpha})^{2}\rangle$ values fall to zero at
different rates, so that the eigendirections associated with high eigenvalues
become locked (i.e., the corresponding velocities become zero) at lower values
of $\beta$ than the ones associated with low eigenvalues. \ Similar trends
were noted for SFN's in \cite{McGrawSpec}. The normal velocites give more
detailed information about synchronization than a global order parameter.
\ The evolution with increasing $\beta$ from complete incoherence to complete
frequency synchronization can be visualized as a series of quasi-independent
locking transitions in which different normal modes effectively drop out of
the active dynamics. \ As eigendirections successively lock, \ the phase space
of the oscillator system can be viewed as contracting onto progressively
lower-dimensional subspaces spanned by the remaining eigenvectors. \ Modes
with very low eigenvalues are very difficult to lock, and therefore their
presence will inhibit complete synchronization. \ Evidently these low modes
are correlated with the difficulty of full synchronization in highly clustered
networks. \ 

\begin{figure}
[ptb]
\begin{center}
\includegraphics[
trim=0.000000in 0.000000in 0.037127in 0.000000in,
height=2.5374in,
width=3.039in
]%
{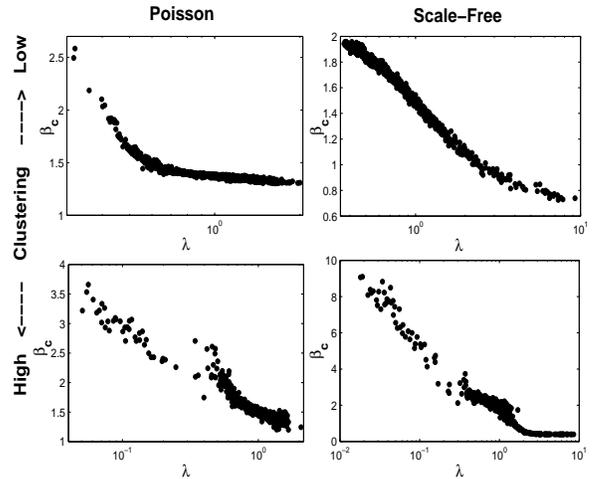}%
\caption{Transition points $\beta_{c}$ for individual eigenvectors of
unnormalized networks. \ $\beta_{c}$ is the coupling strength at which each
mode locks as defined by the criterion $\langle(\Omega^{\alpha})^{2}%
\rangle<0.01$. \ Eigenvalues are plotted on a logarithmic scale in order to
reveal the trend at low eigenvalues. \ \ }%
\label{transitionsU}%
\end{center}
\end{figure}

\begin{figure}
[ptbptb]
\begin{center}
\includegraphics[
trim=0.000000in 0.000000in -0.054172in 0.000000in,
height=2.5365in,
width=3.039in
]%
{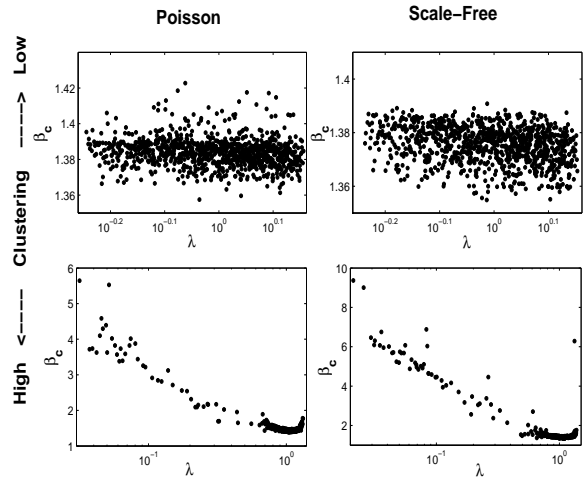}%
\caption{Mode transition points $\beta_{c}$ for normalized networks. \ }%
\label{transitionsN}%
\end{center}
\end{figure}

To the extent that the modes behave as independent degrees of freedom, one can
view each mode as having its own individual transition point, \ a critical
coupling strength above which that mode is locked. \ To obtain numerical
estimates of these individual transitions, we considered a mode to be locked
when the value of $\langle(\Omega^{\alpha})^{2}\rangle$ falls below a
threshold of 0.01. \ \ (Since we are measuring an average over frequency
realizations, this means that for the majority of realizations the value of
$\Omega^{\alpha}$ is actually zero within the resolution of our numerical
measurement.) \ Frequencies were measured at a sequence of values of $\beta$
for one network of each of the types we studied. \ As in figure
\ref{partialdisorder} \ The frequencies $\Omega^{\alpha}$ were time averaged
over 300 time units, and their squares were averaged over 100 realizations of
the random intrinsic frequencies. \ The transition points $\beta_{c}$ were
estimated by means of a cubic spline interpolation of the numerical
measurements. \ \ The results are plotted in figures \ref{transitionsU} and
\ref{transitionsN}. \ The patterns are somewhat different for each type of
network, but as a general rule $\beta_{c}$ is a decreasing function of the
eigenvalue $\lambda$. \ \ In some of the networks, \ a finite subset of the
modes all lock at the same $\beta$ value while the remainder lock and unlock
independently. \ Evidently, some modes are strongly mutually coupled while
others are more independent. \ For the strongly clustered SFN, for example, it
is evident from fig. \ref{transitionsU} that a number of high modes lock
simultaneously at a quite low value of $\beta=0.5$. \ This group of modes
apparently represent the degrees of freedom responsible for the advanced
partial synchronization of this network and the upward turn in the global
order parameter in fig. \ref{orderparam}. \ \ For the normalized networks,
\ all modes within the main spectral peak lock almost simultaneously, \ and
only the lower set of modes (in highly clustered networks) have significant
spread in their values of $\beta_{c}$. \ The low-eigenvalue modes that form at
high clustering are difficult to lock, and it is evidently these modes that
are responsible for the inhibition of complete synchronization in the cases of
highly clustered neworks. \ \ Since these modes are associated with the
divisions among communities, \ this suggests that the frequency clusters noted
at moderately high $\beta$ in \cite{McGraw} are identical to topological
communities. \ The degrees of freedom that inhibit full synchronization are
clearly different from the ones responsible for the advanced partial
synchronization of SFN's. \ \ \

\section{Conclusions\label{Conclusions}}

We have examined the Laplacian spectra for several types of complex networks,
\ seeing the effects on the spectra of degree distribution, clustering, and of
coupling scheme (in particular, equal and bidirectional couplings versus input
normalization.) \ We found that increasing clustering has two main groups of
effects on the eigenvalue spectrum. \ The first set of effects alter the shape
and composition of the main spectral peak. \ More localized modes are formed
in and around this peak, and the correlation between eigenvalue and degree
expectation value generally becomes stronger. \ Second, \ increased clustering
creates an additional group of delocalized low-eigenvalue modes which are
associated with an increased modularity (community structure) of the network.
\ \ \ As expected, \ input normalization greatly reduces the influence of
degree heterogeneity \ compared to the unnormalized network, \ and this is
reflected in a greater uniformity of the spectra. \ Normalization, however,
does not significantly alter the effect of clustering on communities and the
low modes associated with them. \ \ In general, \ study of the Laplacian
spectra and the properties of the eigenvectors reveals complex structures
whose significance should be explored further in future work. \ Many
qualitative features are independent of the average degree $<k>$ but depend
only on the type of topology and coupling scheme. \ \ Some peculiar details of
the spectral shapes we have observed remain to be fully explained \ (see, for
example, figs. \ref{iprcomb} and \ref{devncomb}.) \   

We found that when coordinates based on the Laplacian eigenbasis are used to
examine the dynamics of a network of coupled phase oscillators, \ the
transition to synchronization can be visualized in ways that are not apparent
from global order parameters alone. \ \ Roughly speaking, extremal eigenvalues
give information about the onset of synchronization or desynchronization,
whereas the full spectrum is relevant to the full process of synchronization.
\ \ In particular, in many cases, modes or groups of modes make
quasi-independent transitions to synchronization as the coupling strength is
increased, with low modes synchronizing at higher coupling strengths. \ The
process of synchronization can be viewed as a contraction of the dynamics onto
progressively lower-dimensional submanifolds of the phase space as different
eigenmodes lock one by one. \ The presence of low-lying modes, which are hard
to lock, can significantly retard the achievement of full synchronization.
\ \ The quasi-independence of eigenvectors is a somewhat surprising result in
a highly nonlinear regime of partial synchronization. \ Previously the
Laplacian was applied only to the linear stability of a perfectly synchronized
state. \ 

Unlike the situation with the linear (MSF) problem, \ in the regime of partial
synchronization a knowledge of the eigenvalue spectrum alone is not sufficient
to predict the full dynamics. \ \ The dependence of transition point on
eigenvalue is different for each type of network even though the broad trend
(the transition point is a decreasing function of eigenvalue) is the same for
all. \ But the spectrum is nonetheless a source of at least heuristic insight
into the dynamics, and particular groups of eigenvectors can be directly
associated with aspects of the dynamics: \ for example, high modes with the
advanced transition in high-clustering SFN's, \ or low modes with frequency
clustering and the inhibition of full synchronization.

In the future, diagnostic techniques based on the Laplacian spectrum can be
applied to other types of networks. \ It may be worthwhile to examine spectral
effects of assortative mixing, and to test our conjecture that assortative
mixing should bring the Laplacian and adjacency matrix spectra into closer
congruence. \ It may also be interesting to attempt a finite size scaling
analyis of the separate transitions undergone by different normal modes. \
\ Just as one gains important information by considering the full process of
synchronization and not only the onset\cite{GomezMorenoIJBC}, \ one also gains
information by examining the full spectrum of eigenvectors. \ 

\begin{acknowledgement}
This work was supported by the NSERC of Canada.
\end{acknowledgement}

\end{document}